\documentstyle[preprint,aps,epsfig,psfig]{revtex}

\begin{document}
\title{{\bf A Relativistic Separable Potential to Describe Pairing in Nuclear 
 Matter}}
\author{B. Funke Haas, T. Frederico, B. V. Carlson}
\address{
Departamento de F\'{\i}sica, Instituto Tecnol\'ogico da Aeron\'autica -- 
CTA,\\
12.228-900 S\~ao Jos\'e dos Campos, S\~ao Paulo, Brazil}
\author{F.B. Guimar\~aes}
\address{EAN -- Instituto de Estudos Avan\c cados -- CTA,\\
12.228-840 S\~ao Jos\'e dos Campos, S\~ao Paulo, Brazil}
\maketitle

\begin{abstract}
Using the Dirac-Hartree-Fock-Bogoliubov approximation to study nuclear
pairing, we have found the short-range correlations of the Dirac $^1$S$_0$ pairing
fields to be essentially identical to those of the two-nucleon virtual state
at all values of the baryon density. We make use of this fact to develop a
relativistic separable potential that correctly describes the pairing fields.
\end{abstract}


\section{Introduction}

We have recently studied nuclear pairing in meson-exchange models of the
nuclear interaction using a Dirac-Hartree-Fock-Bogoliubov (DHFB)
approximation to nuclear matter\cite{gcf,cfg}. An important conclusion of
this study is that the short-range $^1$S$_0$ pairing correlations in
nuclear matter are
essentially identical to the short-range correlations of the two-nucleon
virtual state, due to the dominance of the virtual state in the $^1$S$_0$
channel. In non-relativistic calculations, Khodel, Khodel and Clark\cite{kkc}
also noted a close similarity between the pairing gap function and the
short-range vertex function of the virtual state\cite{cms}, which
prompted them to suggest that the virtual state be used as the starting
point in calculating the gap function. In our relativistic calculations,
however, the similarity of the Dirac pairing fields and the vertex functions
is so striking that we can claim that the two are
essentially identical. We make use of this quasi-identity to develop a
separable approximation to the relativistic pairing potential that correctly
takes into account the high momentum contributions of the short-range
two-nucleon correlations.

Another indication that a separable potential can describe nuclear
pairing correlations comes from the usual assumption of a constant pairing gap
in finite nucleus calculations. The analysis of the HFB equations in the
zero-range limit of the two-nucleon interaction shows that the separable
potential approximation becomes an {\em exact} formulation of the pairing
problem in the zero-range limit as the pairing potential then
becomes independent of the baryon momentum. The pairing field also become
independent of the baryon momentum and vary
only with respect to the baryon density of the nuclear matter, which is
equivalent to a constant pairing gap assumption. Thus, the separable
potential approximation for the pairing field can be seen as the simplest
extension, to finite-range two-nucleon interactions and all baryon
densities, of the zero-range limit of the HFB-equations. It gives the
simplest improvement to the usual assumption of a constant pairing gap.

The aim of this paper is to present a rank-one, separable interaction that 
accurately describes the DHFB pairing field at all densities. In section II,
we clarify the close relationship between the pairing field and the 
two-nucleon $^1$S$_0$ virtual state vertex function and use this relationship
to define the separable interaction. In Section III, we demonstrate the 
goodness of the separable interaction through comparisons with exact
DHFB calculations of nuclear matter.

\section{The formalism}

The bound-state correlations in a two-particle system can be roughly
classified as asymptotic ones and short-range ones. The asymptotic
correlations are determined principally by the binding energy, while the 
short-range ones
depend on the high-momentum components of the wave function. This can
be seen by examining the manner in which a bound pair appears in the
two-body T-matrix, $T(E)$. The T-matrix satisfies the integral equation 
\begin{equation}
T(E)=V+V\,G_{0}(E)\,T(E)\,, \label{lipsch}
\end{equation}
where $V$ is the two-body interaction and $G_{0}(E)$ the free two-body
propagator. A bound state corresponds to a pole in the T-matrix at a negative 
value of the energy, $-\epsilon_b$. We can decompose the T-matrix, in this
case, as
\begin{equation}
T(E)=\Gamma\frac{1}{-\epsilon_b -E}\Gamma^{\dagger} + T_c(E)\,,
\end{equation}
where $\Gamma$ is the bound-state vertex function and $T_c(E)$ is the continuum
component of the T-matrix (and, possibly, the contributions of other bound
state poles). Substituting the decomposition in the integral equation,
Eq.~(\ref{lipsch}),
we verify that the vertex function satisfies the equation 
\begin{equation}
\Gamma = VG_0(-\epsilon_b)\Gamma\,.
\end{equation}
The vertex function of the two-nucleon bound state is closely related to its
wave function, $\psi=G_0(-\epsilon_b)\Gamma$, which satisfies the differential
equation
\begin{equation}
(-\epsilon_b-H_0)\psi = (-\epsilon_b-H_0) G_0(-\epsilon_b) \Gamma = \Gamma
=VG_{0}(-\epsilon_b)\Gamma = V \psi\,,
\end{equation}
where $H_0$ is the Hamiltonian of the free two-nucleon system.

Here we see the rough division of the two-nucleon correlations in the
vacuum.  The asymptotic correlations are determined by the singularity
of the Green's function $G_{0}(-\epsilon _{b})$ while the short-range
ones are contained in the high momentum components of the vertex
function, $\Gamma $.  This analysis can also be extended to the
nucleon pairing correlations in nuclear matter. One can directly
associate the bound-state vertex function $\Gamma $ of the T-matrix
with the pairing field of the HFB approximation, $\Delta $, and the
adjoint vertex function, $\Gamma^{\dagger}$, with the time-reverse
conjugate pairing field, $\bar{\Delta}$. To demonstrate this, we
develop the relation between $\Gamma $ and $\Delta $ in a formal way,
by comparing the equations that define the two. (In a similar manner,
the pair wave function $\psi$ can be associated with the residue of
the anomalous propagator $F$ in the complex energy
plane\cite{cfg,h-j}. However, we will not develop this association
here.)

In the Dirac-Hartree-Fock-Bogoliubov (DHFB) approximation to pairing in
nuclear matter, the self-consistency equation for the pairing field can
be written as\cite{gcf,cfg}
\begin{equation}
\Delta (k)= i \sum_j\int \frac{d^{4}q}{(2\pi )^{4}}\Lambda _{j\alpha
}D_{j}^{\alpha \beta }(k-q)F(q)B\Lambda _{j\beta }^{T}B^{\dagger }\,,
\label{del1}
\end{equation}
where, $\Lambda_{j\alpha }$ represents the meson-baryon vertex of meson $j$ 
(for the $\sigma$ meson,  $\Lambda_{\sigma\alpha }=ig_s\openone$, for the
$\omega$ meson, $\Lambda_{\omega\alpha }=-ig_v\gamma_{\alpha}$, etc.), 
$D_{j}^{\alpha\beta }$ represents the propagator of meson $j$ and the Greek
letters represent any necessary Lorentz and/or isospin indices. The matrix $B$
relates transposed quantities to the complex conjugates of time-reversed ones.
It is given by $B=\tau_2\otimes\gamma_5 C$, where the Pauli matrix $\tau_2$
acts in the isospin space and $C$ is the charge conjugation matrix.

The anomalous propagator $F(q)$ in the self-consistency equation, 
Eq.~(\ref{del1}), is one of the number-non-conserving components of the 
baryon propagator $S_F(q)$, which, in the Gorkov formalism, takes the form
\begin{equation}
S_{F}(q) = \left( 
\begin{array}{ll}
G(q) & F(q) \\ 
\widetilde{F}(q) & \widetilde{G}(q)
\end{array}
\right)\,, \label{GF}
\end{equation}
in which $G(q)$ is the usual number-conserving propagator and 
$\widetilde{F}(q)$ and $\widetilde{G}(q)$ are the corresponding 
time-reversed propagators. The baryon propagator $S_F(q)$ can be
written in terms of the self-energy and pairing fields, $\Sigma(q)$ and
$\Delta(q)$, respectively, as
\begin{equation}
S_{F}(q)=\left( 
\begin{array}{cc}
\not{q}-M-\Sigma (q)+\mu \gamma _{0} & \Delta (q) \\ 
\overline{\Delta }(q) & \not{q}+M+\Sigma (q)-\mu \gamma _{0}
\end{array}
\right) ^{-1}\,,  \label{Inversa}
\end{equation}
where $\mu$ is a Lagrange multiplier used to fix the average baryon density.

The form of the $^{1}$S$_{0}$ pairing field can be determined from the
hermiticity and antisymmetry properties of the Lagrangian density, as well
as the translational, rotational and isospin symmetries of symmetric nuclear
matter, as is shown in Ref. \onlinecite{gcf}. The general form obtained for
the field is
\begin{equation}
\Delta (k)=(\Delta _{S}(k)-\gamma _{0}\,\Delta _{0}(k)-\gamma _{0}
\vec{\gamma}\cdot \hat{k}\,\Delta _{T}(k))\vec{\tau}\cdot \hat{n}\,.
\label{deltaform}
\end{equation}
where $k=(k_{0},\vec{k})$ is the baryon 4-momentum, 
$\hat{k}=\vec{k}/|\vec{k}|$ and
the isospin orientation $\hat{n}$ is arbitrary. We note that
$\vec{\tau}\cdot \hat{n}=\tau _{2}$ corresponds to the standard case of 
proton-proton and neutron-neutron pairing.

We can decompose the self-consistency equation for the pairing field, 
Eq.~(\ref{del1}), into equations for the components of Eq.~(\ref{deltaform}).
After integrating over energy and angle and neglecting retardation effects,
the component equations take the form, 
\begin{eqnarray}
\Delta _{S}(k,k_{F}) &=&\frac{1}{2\pi ^{2}}\int_{0}^{\Lambda
}q^{2}dqV_{S}(k,q)F_{S}(q,k_{F})\,,  \nonumber \\
\Delta _{0}(k,k_{F}) &=&\frac{1}{2\pi ^{2}}\int_{0}^{\Lambda
}q^{2}dqV_{0}(k,q)F_{0}(q,k_{F})\,,  \label{paireq} \\
\Delta _{T}(k,k_{F}) &=&\frac{1}{2\pi ^{2}}\int_{0}^{\Lambda
}q^{2}dqV_{T}(k,q)F_{T}(q,k_{F})\,,  \nonumber
\end{eqnarray}
where $k=|\vec{k}|$ and $q=|\vec{q}|$. The pairing potentials, $V_{S}$, 
$V_{0}$ and $V_{T}$ (given in the Appendix), are functions of the coupling
constants and meson masses, while $\Lambda $ is a cutoff in the baryon
momentum. The Fermi momentum $k_F$ is defined through its conventional
relation to the 
baryon density, that is, $\rho _{B}=\gamma k_{F}^{3}/3\pi ^{2}$, where 
$\gamma $ is 2 for nuclear matter and 1 for neutron matter. The components of
the anomalous propagator are defined as
\begin{eqnarray}
F_S(k) &=& \frac{1}{8}\mbox{Tr}[\vec{\tau}\cdot\hat{n}\,F(k)]\,, \nonumber \\
F_0(k) &=& \frac{1}{8}\mbox{Tr}[\gamma_0\vec{\tau}\cdot\hat{n}\,F(k)] \,, \\
F_T(k) &=& -\frac{1}{8}\mbox{Tr}[\gamma_0\vec{\gamma}\cdot\hat{k}
\vec{\tau}\cdot\hat{n}\,F(k)]\,. \nonumber
\end{eqnarray}
These are evaluated in the Appendix in the case in which their negative-energy
Dirac sea contributions are neglected.

The components of the anomalous propagator, $F_S$, $F_0 $ and $F_T $ 
carry the information of the density-dependent nucleon self-energy and
pairing mean-fields. The pairing potentials
$V_S$, $V_0$ and $V_T$ also possess a density dependence associated with
the retardation terms, which account for the finite meson propagation
velocity. In Eqs.(\ref{paireq}), these retardation terms are neglected
since their effects on the pairing fields have been found to be small\cite
{gcf}. The density dependence remaining in the self-consistent pairing
equations, Eqs.(\ref{paireq}), is thus
entirely contained in the components of the anomalous propagator.

When the baryon density $\rho_B$ tends to zero, both the self-energy
$\Sigma(q)$ and the pairing field $\Delta(q)$ also tend to zero.
The anomalous propagator $F(q)$ can then be well approximated by
\begin{equation} 
F(q) \approx - \frac{1}{\not{q}-M+\mu \gamma _{0}+i\eta}
\Delta(q) \frac{1}{\not{q}+M-\mu \gamma _{0}-i\eta}\,,\qquad\qquad
\rho_B\rightarrow 0\,,
\end{equation}
an approximation that becomes exact at zero density. The vacuum limit of the
pairing field thus satisfies the homogeneous equation
\begin{equation}
\Delta (k)=i \sum_j\int \frac{d^{4}q}{(2\pi )^{4}}\Lambda _{j\alpha
}D_{j}^{\alpha \beta }(k-q)\frac{1}{\mu \gamma _{0}+\not{q}-M+i\eta}
\Delta(q) \frac{1}{\mu \gamma _{0}-\not{q}-M+i\eta} 
B\Lambda _{j\beta }^{T} B^{\dagger }\,,\qquad \rho_B\rightarrow 0\,.
\label{del0}
\end{equation}

Turning now to the Bethe-Salpeter equation, we write the bound-state 
vertex function as $\Gamma(k,P)$, where $P$ is the center-of-mass four
momentum of the baryon pair and $k$ is their relative four momentum. The
Bethe-Salpeter equation for the vertex function, in the ladder approximation,
is often written as
\begin{equation}
\Gamma(k,P)=i \sum_j\int \frac{d^{4}q}{(2\pi )^{4}}\Lambda_{1j\alpha}
\Lambda_{2j\beta } D_{j}^{\alpha \beta }(k-q) G_{01}(P/2+q) G_{02}(P/2-q)
\Gamma(q,P)\,, \label{bs}
\end{equation}
where the free baryon propagator $G_0(q)$ is given by
\begin{equation}
G_0(q)=\frac{1}{\not{q}-M+i\eta}\,.
\end{equation}
We can rewrite this in a matrix form, making use
of the matrix B, as 
\begin{equation}
\Gamma(k,P)B^{\dagger}=i \sum_j\int \frac{d^{4}q}{(2\pi )^{4}}\Lambda_{j\alpha}
 D_{j}^{\alpha \beta }(k-q) G_{0}(P/2+q)\Gamma(q,P)B^{\dagger}
 G_{0}(P/2-q) B\Lambda_{j\beta }^T B^{\dagger}\,. \label{bs0}
\end{equation}
Comparing this expression with the self-consistency equation for the vacuum
pairing field, Eq.~(\ref{del0}), we find the two are identical if we
\begin{itemize}
\item{evaluate the vertex function in the center-of-mass frame, taking
\begin{equation}
P_{\alpha}=2\mu\,\delta_{\alpha 0},
\end{equation}
where $\mu$ is the Lagrange multiplier of the Gorkov propagator $S_F(q)$ of
Eq.~(\ref{Inversa}), and}
\item{associate the pairing field and the vertex function as
\begin{equation}
\Delta(k)=\Gamma(k)B^{\dagger}\,,
\end{equation}
where we now suppress the center-of-mass dependence of the vertex function.}
\end{itemize}
We thus conclude that the self-consistency equation for the vacuum pairing
field is identical to the ladder approximation to the Bethe-Salpeter equation
in the center-of-mass frame.

An analysis of the symmetry and antisymmetry properties that we expect of the 
center-of-mass Bethe-Salpeter vertex function, $\Gamma(k)$, 
leads us to the following form,
\begin{equation}
\Gamma (k)B^{\dagger} = \left( \Gamma _{s}(k)-\gamma _{0}\Gamma _{0}(k) 
 -\gamma _{0}\vec{\gamma}\cdot \hat{k}\Gamma _{T}(k)\right) 
\vec{\tau}\cdot \hat{n}\;,
\end{equation}
which is identical to that of the pairing field $\Delta(k)$ given in 
Eq. (\ref{deltaform}). We substitute this vertex function in the
Bethe-Salpeter equation, Eq. (\ref{bs0}), integrate over energy and angle,
and neglect retardation effects and the contribution of the negative energy
states, just as we have done in the self-consistency equation of the pairing
field. We find,
\begin{eqnarray}
\Gamma _{S}(k) &=& \int_{0}^{\Lambda }V_{S}(k,q) \frac{q^{2}}{4\pi^{2}} 
\frac{E_{q} \Gamma _{S} -M \Gamma _{0} }
{E_{q}(M_{B}-2E_{q})} dq\;, \nonumber \\
\Gamma _{0}(k) &=& \int_{0}^{\Lambda }V_{0}(k,q) \frac{q^{2}}{4\pi^{2}}
 \frac{ \left( 2q^{2}-M_{B}E_{q}\right) \Gamma_{0}+MM_{B}\Gamma _{S} }
{M_{B}E_{q}(M_{B}-2E_{q})} dq\;, \label{gammaeq} \\
\Gamma _{T}(k) &=& -\int_{0}^{\Lambda }V_{T}(k,q)\frac{q^{2}k}{4\pi ^{2}}
\frac{q(M_{B}\Gamma _{S}-2M\Gamma _{0})}{M_{B}E_{q}(M_{B}-2E_{q})} dq\;,
\nonumber 
\end{eqnarray}
where $k=|\vec{k}|$ , $q=|\vec{q}|$, $\Lambda$ is a cutoff in the baryon
momentum, and $E_{q}=\sqrt{q^{2}+M^{2}}$. 
The potentials $V_S$, $V_0$, and $V_T$ are, of course, the same as those of the
self-consistency equations for the pairing field, Eqs. (\ref{paireq}),
and are given in the Appendix. The center-of-mass energy of the bound 
two-nucleon pair, that is, the bound two-nucleon mass, $M_B$, 
is given by 
\[
M_{B} = 2\mu = 2M-\epsilon _{B}\,.
\]
In the Appendix, the equivalence between the vacuum pairing equation and the 
ladder approximation to the Bethe-Salpeter equation is once again 
demonstrated, by reducing the explicit form of the components of the
pairing equation, Eqs.~(\ref{paireq}), to the components of the Bethe-Salpeter
equation, Eqs.~(\ref{gammaeq}), in the vacuum limit.

The above expressions are valid for a bound two-body system in the vacuum.
If the pole of the T-matrix corresponds to a virtual state (anti-bound state),
as in the case of the physical $^1$S$_0$ two-nucleon channel,
one must redefine the integrands through an analytic continuation into the
second sheet of the complex energy plane\cite{mcvoy1}. The
analytical continuation introduces an additional term, which takes
into account the discontinuity of the propagator across the cut in the
energy plane. In the components of the Bethe-Salpeter equation, 
Eqs.~(\ref{gammaeq}), in which the negative-energy states have been neglected, 
the analytically continued equations can be obtained through the substitution,
\begin{equation}
\frac{1}{M_{B}-2E_{q} } \rightarrow \frac{1}{M_{v}-2E_{q} } -2\pi
i\delta (M_{v}-2E_{q})\,,
\label{denom3}
\end{equation}
where $M_{v}$ is the mass of the two-nucleon virtual state,
\[
M_v=2M-\epsilon_v=\sqrt{M^{2}-|q_{v}|^{2}}\,, 
\]
with $q_{v}=-i|q_{v}|$.

The equations for the components of the vertex-function then become 
\begin{eqnarray}
\Gamma _{S}(k) &=& \Gamma _{S}(k)_{v}-\frac{|q_{v}|}{\pi }\left\{
V_{s}(k,q)\left(E_q \Gamma _{S} + M \Gamma _{0}\right) \right\}
_{q=q_{v}}\,, \nonumber \\
\Gamma _{0}(k) &=& \Gamma _{0}(k)_{v}+\frac{|q_{v}|}{\pi }\left\{
V_{0}(k,q) \frac{\left( 2q^{2}-M_{v}E_{q}\right) \Gamma
_{0}-MM_{v}\Gamma _{s} }{M_{v}} \right\} _{q=q_{v}}\,, 
\label{gamma} \\
\Gamma _{T}(k) &=& \Gamma _{T}(k)_{v}+\frac{|q_{v}|}{\pi }\left\{
kV_{T}(k,q)\frac{q(M_{v}\Gamma _{S}+2M\Gamma _{0})}{M_{v}}
\right\} _{q=q_{v}}\,.  
\end{eqnarray}
The functions $\Gamma _{s}(k)_{v},\;\Gamma _{0}(k)_{v}$ and $\Gamma
_{T}(k)_{v}$ are given by Eqs.(\ref{gammaeq}) with the mass of the bound state
$M_B$ substituted by that of the virtual state $M_v$.

\begin{figure}[tbp]
\begin{center}
\epsfig{file=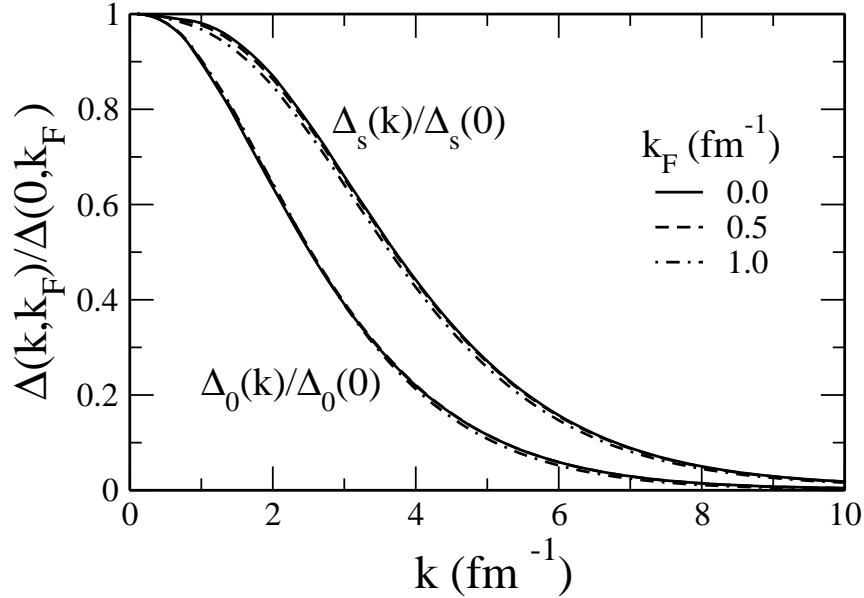,height=8cm}
\caption{The momentum dependence of the two large components of the Dirac
pairing field, $\Delta_{S}$ and $\Delta_{0}$, are shown at several values of the nuclear
matter density, using the  parameters of Ref. [7]. The curve
labeled as zero density corresponds to the vertex function of the virtual state.}
\label{fig1}
\end{center}     
\end{figure}

As stated in the introduction, we have found that the momentum
dependence of the components of the pairing field $\Delta$,
calculated with various sets of interaction parameters, shows a very
small variation over a wide range of values of the baryon density and a 
wide range of cut-offs ($\Lambda $ between 2.5 fm$^{-1}$ and 15.0 
fm$^{-1}$)\cite{cfg}. As an example, in Fig. \ref{fig1} we show the momentum 
dependence of the principal components of the pairing field, $\Delta_S(k)$ and
$\Delta_0(k)$, for several values of the Fermi momentum, which were obtained
using the $\sigma-\omega$ parameters of Ref.\cite{bouyssy} with a momentum
cutoff of $\Lambda=$ 10 fm$^{-1}$. Note that the curves denoted by Fermi momentum
zero correspond to the components of the vertex function of the 
$^1$S$_0$ virtual state in the vacuum. The small variation in the 
components as a function of the Fermi momentum suggests the use of the
momentum dependence at one value of the density. Use of the vacuum momentum
dependence is particularly convenient, since in this case a direct
correspondence exists between the self-consistent pairing field and
the Bethe-Salpeter vertex function. A direct comparison between the numerical
solutions for the vertex function, $\Gamma (k)$, and for the pairing field,
$\Delta(k)$, gives us the following very good approximate relations between 
the components of the two fields, 
\begin{eqnarray}
\Delta _{S}(k,k_{F}) &\approx& d(k_{F})\,\Gamma _{S}(k)\,,  \nonumber \\
\Delta _{0}(k,k_{F}) &\approx& d(k_{F})\,\Gamma _{0}(k)\,,\ \label{eq:s11} \\
\Delta _{T}(k,k_{F}) &\approx& d(k_{F})\,\Gamma _{T}(k)\,, \nonumber
\end{eqnarray}
where $d(k_{F})$ is a density-dependent factor determining the overall
magnitude of the pairing fields. 

We emphasize that, in the nonrelativistic approach of Khodel, Khodel and Clark,
the virtual state vertex function serves only as a good starting point for
calculating the pairing field\cite{kkc}, while in our relativistic approach,
the two are essentially identical in form. We can explain the difference
between the two approaches by analyzing the effective gap function for
the positive energy DHFB states, which takes the form
\begin{equation}
\Delta _{g}(k)=\Delta _{0}\frac{M^{\star }}{E_{k}^{\star }}-\Delta
_{S}+\Delta _{T}\frac{k^{\star }}{E_{k}^{\star }}\,,
\end{equation}
where $E^{\star }=\sqrt{k^{\star 2}+M^{\star 2}}$ and the effective momentum
and mass of the quasi-nucleons, $k^{\star}$ and $M^{\star}$, are defined in
the Appendix. The gap function $\Delta _{g}(k)$ is the quantity whose
role is closest to that of the non-relativistic pairing field\cite{gcf}.
Like the non-relativistic self-energy, the gap function is the
difference between two larger relativistic quantities, $\Delta _{0}$ and 
$\Delta _{S}$. ($\Delta _{T}$ is two to three orders of magnitude smaller
than the other components.) The effective mass $M^*$ that appears in the
gap function varies strongly with the nuclear matter density. Although
the form of the components of the Dirac pairing field remain almost
constant with density, the gap function changes, due to its dependence
on $M^*$. The density independence of the form of the relativistic pairing
field thus does not carry over to the nonrelativistic field.

The invariant momentum dependence of the components of the pairing fields
suggests a separable form for an effective potential, in which each of the 
components of the interaction, $V_S$, $V_0$ and $V_T$, is proportional to
the corresponding component of the vertex function, $\Gamma_S$, $\Gamma_0$,
and $\Gamma_T$. Such potentials are well known from the 
nonrelativistic treatment of the two-nucleon problem. It is such a
separable form of the non-relativistic pairing potential that Khodel, Khodel
and Clark suggest as a starting point for calculating the nonrelativistic gap
function\cite{kkc}. Here, we use a form appropriate for the components of the
relativistic pairing field that also preserves the symmetry in the two 
momenta, $k$ and $q$, of the potential, taking
\begin{eqnarray}
V_{S}(k,q) &=& \frac{\lambda _{S}}{4}\Gamma _{S}(k)\Gamma _{S}(q)\;,
\nonumber \\
V_{0}(k,q) &=& \frac{\lambda _{0}}{4}\Gamma _{0}(k)\Gamma _{0}(q)\;,
\label{eq:s8} \\
kV_{T}(k,q) &=& \frac{\lambda _{T}}{4}\Gamma _{T}(k)\Gamma _{T}(q)\;,
\nonumber
\end{eqnarray}
where the potential strengths, $\lambda _{S}$, $\lambda _{0}$, and
$\lambda_{T}$, are obtained through the simultaneous solution of
Eqs.(\ref{gammaeq}) and (\ref{eq:s8}), in the case of a bound state,
or Eqs.(\ref{gamma}) and (\ref{eq:s8}), in the case of a virtual state. 
We note that, although we have included the equation for the tensor component
$\Gamma _{T}$ in Eqs. (\ref{eq:s11}) and (\ref{eq:s8}), we have neglected
this component in our numerical calculations. The inclusion of the tensor 
term offers little further calculational difficulty but, being
extremely small, has little effect on the rest of the calculation.

For any given meson-exchange potential, we have shown in Ref.\cite{cfg}
that the size of the pairing gap, for low
baryon densities, is correlated with the energy of the virtual state of the
T-matrix, $\epsilon_v $. The value of $\epsilon_v $, in turn is related to 
the baryon momentum cut-off, $\Lambda $, in the integrals of the
self-consistency equations. Consequently, fixing $\epsilon_v $ at its
physical value in the Bethe-Salpeter equation for $\Gamma$ determines a
(usually unique) value for $\Lambda $.

The separable potential introduces two additional parameters, $\lambda _{S}$
and $\lambda _{0}$, when the tensor term is neglected. These parameters,
together with the momentum cutoff, $\Lambda$, are
fixed by requiring that the energy of the two nucleon state and the ratio
of the vertex functions be equal to the respective ones obtained by
solving the Bethe-Salpeter equation with a meson-exchange potential.
Together, the parameters $\lambda _{S}$, $\lambda _{0}$ and $\Lambda$ 
introduce an ambiguity in the determination of the separable potential.
Our calculations have shown that, within a wide range of values of the
cutoff $\Lambda $, we can find values of $\lambda _{S}(\Lambda)$ and
$\lambda _{0}(\Lambda)$
that fix the virtual state energy and the vertex functions at their physical
values. We regard this as a positive feature of the separable potential, as it 
permits us to fix one of the parameters, usually $\Lambda$, at an arbitrary, 
and possibly convenient, value, and then solve for the others.  

The only task remaining to complete the definition of the separable pairing
potential is to extend the potential, which we have defined in the
vacuum, to finite values of the baryon density. This is done by observing
that, as the retardation terms in the pairing potentials have been
neglected, the dependence of the pairing field on the baryon density is
completely determined by the density dependence of the components of the
anomalous density, $F_S$, $F_0$
and $F_T $, in Eq.(\ref{paireq}). Consequently, the separable pairing
potential for finite baryon densities must be the same as the potential
in the vacuum. The self-consistency equations for the components of the
pairing field are thus given by Eqs. (\ref{paireq}) and (\ref{eq:s8}),
with the parameters $\lambda_S$, $\lambda_0$, and $\Lambda$ (and, eventually,
$\lambda_T$ as well) fixed by the vacuum solution. 

\section{Results and conclusion}

We have compared calculations of the pairing fields obtained using a separable
interaction with the usual DHFB results for various sets of $\sigma
-\omega $ interaction parameters. The relativistic separable potential
describes well both the gap parameter, $\Delta _{g}$ , and the two large
components, $\Delta _{0}$ and $\Delta _{s}$, at all baryon densities. It
also reproduces the behavior of the pairing field $\Delta $ for increasing
values of the momentum cut-off $\Lambda $ in the self-consistency equations.

\begin{figure}[tbp]
\begin{center}
\epsfig{file=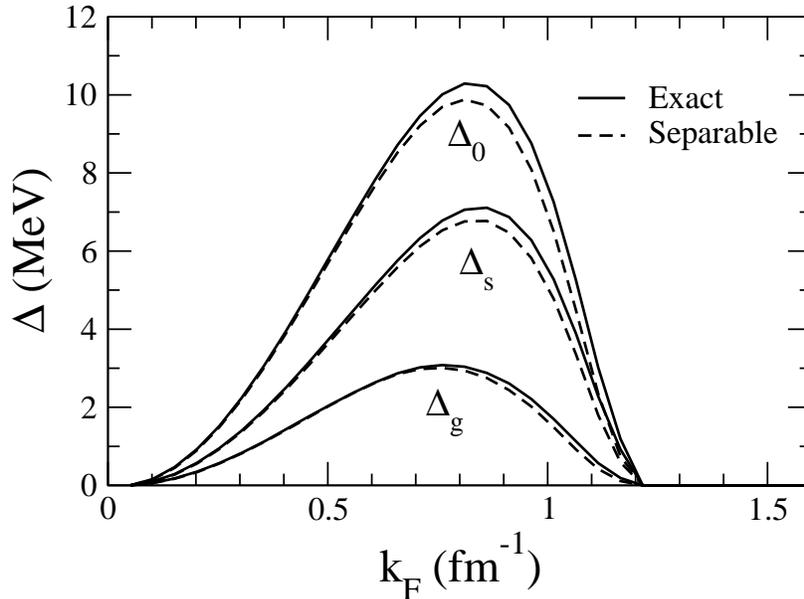,height=8cm}
\caption{The pairing fields obtained in exact and separable DHFB
calculations, using the $\sigma-\omega$ parameters of Ref. [7], are
shown. The two large components of the Dirac pairing field, $\Delta_S$ and 
$\Delta_0$, and the gap parameter, $\Delta_g $, are shown as functions of the
baryon density, for $\Lambda =$ 3.8 fm$^{-1}$.}
\label{fig2}
\end{center} 
\end{figure}

In Fig. \ref{fig2} we show the particular results for the $\sigma-\omega$
interaction of Bouyssy {\em et.al.}\cite{bouyssy}, for a cut-off of $\Lambda
=$ 3.8 fm$^{-1}$, obtained by adjusting the virtual state energy to its
physical value. This potential has been chosen because its vacuum gap function,
which is the nonrelativistic two-nucleon vertex function, is in good agreement
with the nonrelativistic two-nucleon vertex function calculated with the Bonn-B
potential\cite{mach3}. We note a difference of at most a few percent between 
the components of the relativistic fields obtained in exact and separable
calculations. The differences in the two gap functions are even smaller. 
Similar results are obtained for other sets of $\sigma-\omega$ interaction
parameters\cite{cfg}.

\begin{figure}[tbp]
\begin{center}
\epsfig{file=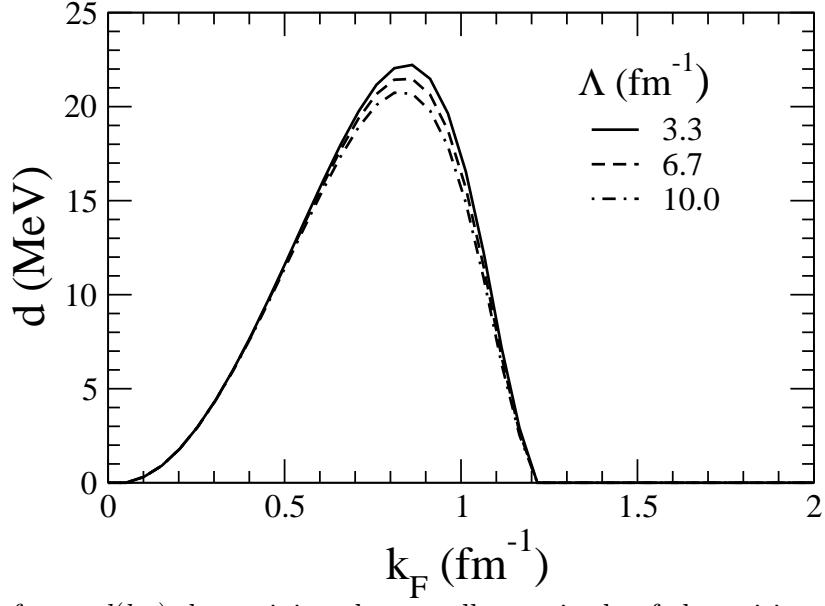,height=8cm}
\caption{The factor $d(k_F)$ determining the overall magnitude of the pairing 
fields, calculated using the $\sigma-\omega$ parameters of Ref. [7],
is shown for several values of the momentum cutoff $\Lambda$.}
\label{fig3}
\end{center}            
\end{figure}

As discusses above, it is possible to adjust the strengths $\lambda_{S}$
and $\lambda_{0}$ 
of the separable interaction to fit the experimental value of the two-nucleon
virtual state energy as a function of the cutoff $\Lambda$. For each value of
the $\Lambda$, the resulting values of the strengths $\lambda_S$ and
$\lambda_0$
yield a separable potential which carries the physical information necessary
for the correct evaluation of the gap parameter at all values of the density. 
Since the properties of the virtual state dominate the physics of $^1$S$_0$
pairing, we would expect the pairing fields that result to be independent of 
the cutoff used. We can verify the extent to which this is true by comparing
the results obtained for $d(k_F)$ of Eq. (\ref{eq:s11}), the common factor
determining the overall
magnitude of the pairing fields, at different values of the momentum cutoff 
$\Lambda$. We do this in Fig. \ref{fig3}, where we display the factor 
$d(k_F)$ as a function of the Fermi momentum for several values of the cutoff
$\Lambda$. We find the form of the factor $d(k_f)$ to be fairly constant but
its magnitude to decrease between 5 and 10 \% as the momentum cutoff $\Lambda$
increases from 3.3 fm$^{-1}$ to 10 fm$^{-1}$.

\begin{figure} [tbp]
\begin{center}
\epsfig{file=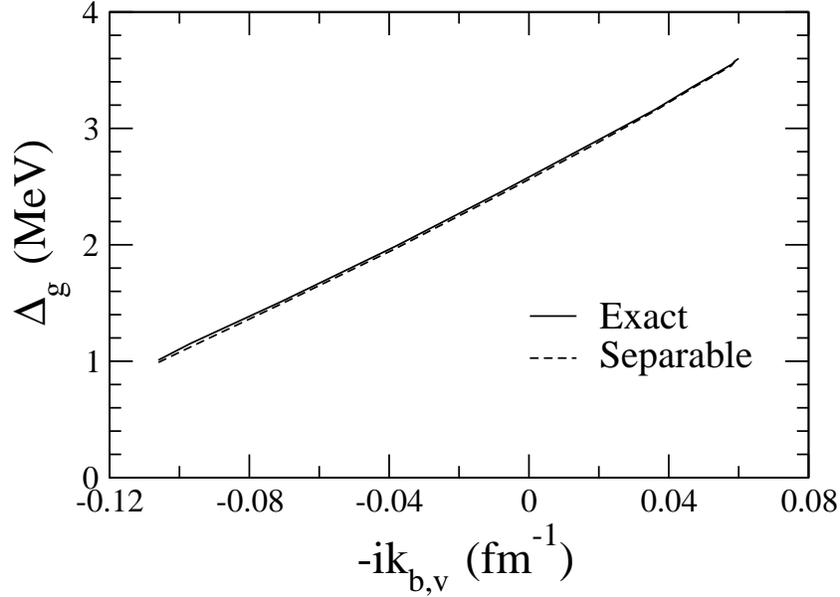,height=8cm}
\caption{The magnitude of the pairing gap as a function of the 
energy-equivalent momentum of the two-nucleon virtual state from exact and 
separable DHFB calculations, using the $\sigma-\omega$ parameters of
Ref. [7], are shown. The positive and negative
values of the abscissa correspond to bound and virtual two-nucleon states,
respectively.}
\label{fig4}
\end{center} 
\end{figure}

In Ref.~\cite{cfg}, a strong correlation between the magnitude of the 
pairing gap and the energy of the two-nucleon virtual state was observed
in all the meson potentials studied there. The correlation is reproduced,
for all values of the cut-off, when using the corresponding separable 
potential. In Fig. \ref{fig4}, we compare the correlation for the gap
parameter, $\Delta_{g}$, obtained with
the separable interaction with the exact result, using the
$\sigma-\omega$ parameters of Ref.\cite {bouyssy}. The agreement between the 
two is excellent, confirming once again the goodness of the separable
approximation to the pairing interaction.

We thus conclude that a separable interaction obtained by exploiting the
close relationship between the Dirac pairing field and the virtual state 
vertex function furnishes an excellent description of the Dirac pairing
fields at all values of the baryon density. We next plan to implement a
configuration-space version of the separable
interaction in a DHFB code for finite nuclei calculations. We expect that
such a potential will describe the short-range pairing correlations with
greater accuracy than those obtained with the usual constant gap assumption.

\vspace{2.5truecm}

\section*{Acknowledgments}

B.V. Carlson and T. Frederico acknowledge the partial support of the
Funda\c{c}\~{a}o de Amparo \`{a} Pesquisa do Estado de S\~{a}o Paulo -
FAPESP and of the Conselho Nacional de Desenvolvimento Cient\'{\i}fico e
Tecnol\'ogico - CNPq, Brazil. B. Funke Haas acknowledges the support of the
Funda\c{c}\~{a}o de Amparo \`{a} Pesquisa do Estado de S\~{a}o Paulo -
FAPESP, Brazil.

\appendix

\section*{The Anomalous Propagator and the Pairing Potential}

The functions $F_S , F_0 $ and $F_T $ in Eq.(6) are given by\cite{gcf},

\begin{eqnarray}
F_S(k) = \left\{ \Delta_s [- \alpha + \beta + E_F^{\star 2} + M^{\star 2} +
k^{\star 2} ] + \Delta_{S}^{*} ( \Delta_{S}^2 - \Delta_{0}^2 +
\Delta_{T}^2 k^2 ) \right . \nonumber \\
\left .  - 2M^{\star} E_F^{\star} \Delta_0 
- 2 k^{\star} E_F^{\star} k\Delta_T \right\}
/(4 \beta \omega_{-}) \;,  \nonumber
\end{eqnarray}

\begin{eqnarray}
F_0(k) = \left\{ \Delta_0 ( \alpha - \beta - E_F^{\star 2} - M^{\star 2} +
k^{\star 2} ) + \Delta_{0}^{*} ( \Delta_{S}^2 - \Delta_{0}^2 +
\Delta_{T}^2 k^2 ) \right .  \nonumber \\
\left . + 2 M^{\star} E_F^{\star} \Delta_s - 2 k^{\star} M^{\star} k 
\Delta_T \right\}/(4 \beta \omega_{-}) \;\;,  \label{eq:c14}
\end{eqnarray}
and

\begin{eqnarray}
F_T(k)=\left\{ -k\Delta _{T}(-\alpha +\beta +E_{F}^{\star
2}-M^{\star 2}+k^{\star 2})-k\Delta _{T}^{*}(\Delta
_{S}^{2}-\Delta _{0}^{2}+\Delta _{T}^{2}k^{2})\right.\nonumber \\
\left. +2k^{\star }E_{F}^{\star }\Delta _{S}-2M^{\star }k^{\star }\Delta
_{0}\right\} /(4 \beta \omega _{-} )\;,  \nonumber
\end{eqnarray}
where the asterisk indicates complex conjugation and $k=|\vec{k}|$. The
quantity $\omega_{-}$ is 
\begin{equation}
\omega _{-}=\sqrt{\alpha -\beta }
\end{equation}
where 
\begin{eqnarray}
\alpha &=& E_{F}^{\star 2}+k^{\star 2}+M^{\star 2}+|\Delta _{S}|^{2}+|\Delta
_{0}|^{2}+|\Delta _{T}|^{2}\;, \nonumber \\ 
\beta &=& 2\sqrt{\beta _{1}^{2}+\beta _{2}^{2}+\beta _{3}^{2}+\beta _{4}^{2}}
\;, \nonumber \\
\beta _{1} &=& -\mbox{Im}(\Delta _{T}\Delta _{0}^{* })\;, \label{bets} \\
\beta _{2} &=& E_{F}^{\star }M^{\star }+\mbox{Re}(\Delta _{S}
\Delta _{0}^{*})\;, \nonumber \\ 
\beta _{3} &=& E_{F}^{\star }k^{\star }+\mbox{Re}(\Delta _{S}
\Delta _{T}^{*})\;, \nonumber \\
\beta _{4} &=& |M^{\star }\Delta _{T}+k^{\star }\Delta _{0}|\;.\nonumber
\end{eqnarray}
The effective momentum and mass of the baryons, $k^{\star}$ and 
$M^{\star}$, are given in terms of the components of the nucleon self-energy,
\begin{equation}
\Sigma(\vec{k})=\Sigma_S(k)-\gamma_0 \Sigma_0(k)-
\vec{\gamma}\cdot\vec{k}\Sigma_V(k)\,,
\end{equation}
as,
\begin{equation}
k^{\star }=(1+\Sigma _{V}(k))k
\qquad\mbox{and}\qquad
\,M^{\star}=M+\Sigma _{s}(k)\,.
\end{equation}
The Fermi energy is 
\[
E_{F}^{\star }=\Sigma _{0}(k)+\mu\,,
\]
where $\mu $ is the chemical potential.

The pairing potentials that take into account the exchange of the mesons 
$\sigma$ and $\omega$ are given by 
\begin{eqnarray}
V_{S}(k,q) &=& \frac{2}{kq}\;\left[ \frac{-1}{8}g_{\sigma }^{2}\theta
_{\sigma}+\frac{1}{2}g_{\omega }^{2}\theta _{\omega}\right]\,,
\nonumber \\
V_{0}(k,q) &=& \frac{2}{kq}\;\left[ \frac{1}{8}g_{\sigma }^{2}\theta
_{\sigma}+\frac{1}{4}g_{\omega }^{2}\theta _{\omega}\right]\,,
\label{eq:vi2} \\
V_{T}(k,q) &=& \frac{1}{2k^{2}q} g_{\sigma }^{2}\phi_{\sigma}  \nonumber
\end{eqnarray}
where the functions $\,\theta _{a}\,$ and $\,\phi _{a}\,$ are 
\begin{equation}
\theta _{a}\;=\;\mbox{ln}\left( \frac{H_{a}+2kq}
{H_{a}-2kq}\right) \;\;,  \label{eq:vi4}
\end{equation}
with 
\begin{equation}
H_{a} = (\omega _{-}(k)-\omega _{-}(q))^{2}-k^{2}-q^{2}-m_{a}^{2}\;, 
\label{eq:vi5}
\end{equation}
and 
\begin{equation}
\phi _{a} = \,1\,-\,\frac{H_{a}}{4kq}\,\theta_{a}\;.  \label{eq:vi6}
\end{equation}
Here $a$ indicates the mesons of the model, with $a = \sigma $, $\omega$. 

In the limit of zero baryon density, we have
\[
k^{\star }\rightarrow k\text{,}\qquad M^{\star }\rightarrow M
\text{, and } \mu\rightarrow E_{F}^{\star }\rightarrow M_B/2\,,
\]
so that
\[
\alpha \rightarrow \frac{M_B^{2}}{4}+E_{k}^{2}\,,\qquad 
\beta \rightarrow M_B E_{k}\,,\qquad \mbox{ and }\qquad
\omega _{-}\rightarrow\frac{1}{2}\left|M_B -2 E_{k}\right|\,, 
\]
where $E_k=\sqrt{k^2+M^2}$. The components of the pairing field tend to zero.
The zero baryon density limit of the pairing self-consistency equations are
then
\begin{eqnarray}
\Delta _{S}(k) &=& \int_{0}^{\Lambda }V_{S}(k,q)\frac{q^{2}}{4\pi^{2}}
 \frac{E_{q} \Delta _{S} -M \Delta _{0}}
{E_{q}(M_{B}-2E_{q})} dq, \nonumber \\
\Delta _{0}(k) &=& \int_{0}^{\Lambda }V_{0}(k,q) \frac{q^{2}}{4\pi^{2}}
\frac{ \left( 2q^{2}-M_{B}E_{q}\right) \Delta_{0}+MM_{B}\Delta_{S} }
{M_{B}E_{q}(M_{B}-2E_{q})} dq\;, \nonumber \\ 
\Delta _{T}(k) &=& -\int_{0}^{\Lambda }V_{T}(k,q)\frac{q^{2}k}{4\pi ^{2}}
\frac{q(M_{B}\Delta _{S}-2M\Delta _{0})}{M_{B}E_{q}(M_{B}-2E_{q})} dq\;,
\nonumber 
\end{eqnarray}
which have exactly the same form as the equations for the components of the
vertex function $\Gamma $ given in Eqs.(\ref{gammaeq}).

\end{document}